\documentclass[%
 reprint,
 nofootinbib,
 amsmath,amssymb,
 aps,
]{revtex4-2}

\usepackage{graphicx}
\usepackage{dcolumn}
\usepackage{bm}
\usepackage{relsize}
\usepackage{url}
\usepackage{breakurl}
\usepackage[breaklinks]{hyperref}

\begin{document}


\title{Stripping mechanisms and remediation for H$^-$ beams}

\author{B.T. Folsom}
 \email{ben.folsom@ess.eu}
\author{M. Eshraqi}%
\author{N. Blaskovic Kraljevic}%

\affiliation{%
 European Spallation Source\\
 Lund, Sweden
 }%
\author{B. Gålnander}
\affiliation{Department of Physics and Astronomy, Uppsala University\\ Uppsala, Sweden
}%
\altaffiliation[Also at ]{European Spallation Source;
 Lund, Sweden}


\date{\today}

\begin{abstract}
Negative hydrogen ions are often used for injecting protons from linacs to storage rings via charge-exchange injection. In this process, the two electrons are stripped by a foil or laser to produce protons which can be merged with an existing beam without significantly affecting its dynamics, allowing high intensities of protons to be accumulated. However, this capability comes with the drawback that the outer electron of an H$^-$ ion has a low binding energy and can easily be stripped away prior to injection. This paper addresses the following stripping mechanisms: interactions with residual gas in the beam pipe, blackbody radiation from accelerator components, and electromagnetic fields from accelerator optics (Lorentz-force stripping) and particles within the bunch itself (intrabeam stripping); with a discussion on how to avoid excessive activation from stripped H$^0$ particles and protons. We also demonstrate that the proportion of stripped H$^0$ colliding with a nearby beam pipe or machine-element walls presents only roughly 10\% of those lost in stripping; the remaining stripped particles traverse to the end of a linac or local straight section, which may relax the limits for allowable stripping-based beam loss in H$^-$ accelerators.
\end{abstract}

\maketitle


\section{\label{sec:level1}Introduction}
The H$^-$ ion is particularly useful for applications requiring the accumulation of high-intensity proton populations in a storage ring. These include neutron spallation, pion production for neutrino-physics studies, or injection into high-energy colliders \cite{kustom_overview_2000,Arnaudon:1004186,dracos_essnusb_2020}.  In this context, charge-exchange injection strips H$^-$ ions of both electrons using either thin intercepting foils or laser stimulation~\cite{cousineau_laser_2018}. This entails bringing the H$^-$ ions into orbit with a circulating proton bunch, then converting them to protons without significantly affecting the bunch trajectory. In other words, this process circumvents Liouville's theorem by adiabatically introducing protons into the phase space of a circulating beam.

The drawback of using H$^-$ is its tendency to have its outer electron easily stripped away due to a low binding energy of approximately 0.75~eV~\cite{armstrong_empirical_1963}. This produces an electron and an H$^0$ particle, which carries almost all the energy of the original H$^-$. Since the H$^0$ is neutral, it follows a drift trajectory to ultimately collide and deposit its energy into machine-element walls or a line-of-sight beam dump at the end of a linear section. Such stripping can occur in a number of ways. Loosely ordered from low to high severity in a linac with moderate final energy (1--10~GeV scale), these are: residual gas, blackbody radiation, Lorentz-force or field-induced, and intrabeam stripping (IBSt).

A related phenomenon is double stripping, which typically occurs near the H$^-$ source. This produces protons that can cause excessive activation or structural damage, particularly if they are inadvertently accelerated and propagated as far as, for example, a junction where H$^-$ are diverted away from their linear trajectory toward an accumulator ring or other apparatus~\cite{ikegami_2012_beamcommiss}.

In this work, we summarize the theory governing the various types of stripping, with some discussion on how to avoid excessive activation from stripped H$^0$ particles and protons. Simulated results are shown for practical test cases, where we observe a significant disparity between instantaneous power loss in a beam due to stripping and the resulting power deposition into the beam pipe or other machine elements downstream. The interplay of Lorentz stripping and IBSt (and their overall dependence on beam parameterization) is also quantified.

\section{H$^-$ Stripping Phenomena}
\subsection{Residual-Gas and Double Stripping}
The theory for the stripping of a negative ion by collision with a neutral atom is well established, although systematic experimental data is lacking for energies exceeding ${\sim}$800~MeV. Its resulting fractional loss per unit length $L$ in the lab frame in terms of particle count $N$ can be modeled as \cite{armstrong_empirical_1963,gillespie_high-energy_1977,carneiro_numerical_2009,raparia_double_proton} 

\begin{align}
    &\tau = \sum_i \frac{1}{\rho_i \sigma_i \beta c}
    \nonumber \\
    &\frac{\Delta N}{L} = \frac{1}{\tau \beta c}~~~,
\end{align}
where $\tau$ is the H$^-$ ion lifetime in the presence of residual gas species $i$ with molecular density $\rho_i$ and scattering cross section $\sigma_i$ (e.g. ${\sim}10^{-17} \textrm{cm}^2$ for a 1~MeV H$^-$ beam in H$_2$), and $\beta c$ is the particle velocity.

Since the loss rate is limited for high $\beta$, this type of stripping is often predominant in the low-energy beam transport (LEBT) section adjacent the ion source, where it is standard practice to inject a neutral diatomic gas to compensate for space charge. Typically, the gas pressure required for well-saturated space charge compensation is at least an order of magnitude below where stripping causes substantial beam loss \cite{valerio-lizarraga_negative_2015}. While this should not reach the scale of intrabeam or Lorentz stripping, it can cause a non-negligible fraction of the total stripping \cite{chou_high_2005}. This is discussed further in Sec.~\ref{A2R}.

Double stripping of H$^-$ is also most common in the LEBT (occurring either directly or with H$^0$ as an intermediary); it can comprise a large percentage of the total stripping losses in the case of high neutral-gas pressures. These double-stripped proton trajectories should be simulated for any H$^-$ beamline, especially those requiring high neutral gas pressures for space-charge compensation. Otherwise, a fraction of the protons which traverse coherently can be dumped in a single spot and cause structural damage \cite{raparia_double_proton_fixed}. At J-PARC in Tokai Japan, a chicane was installed in the medium-beta transport, just following the chopper, to divert protons generated in the high-pressure LEBT. This has been effective, although their recommendation is to avoid solenoid focusing in the LEBT (i.e. using a focus--drift--defocus, or FODO, lattice) to prevent proton capture at the outset \cite{ikegami_2012_beamcommiss}.

At higher energies, these double-stripped protons are a concern in linacs where the superconducting RF frequency triples the normal-conducting one, thus capturing protons within accelerating buckets \cite{plum2013challenges,maruta_beamloss_2011}. For linacs where the frequency is doubled between the normal-conducting and superconducting sections, the protons are instead decelerated and, consequently, do not reach problematic energies in terms of activation \cite{plum_beam_2014,plum_beam_2016}.
 
\subsection{\label{BBS} Blackbody Radiation Stripping}
When the infrared spectra of photons emitted from the beam pipe or other beamline elements are Lorentz shifted into the beam's reference frame, they can reach sufficient energy for H$^-$ stripping. 
\begin{figure}
  \includegraphics[width=0.48\textwidth]{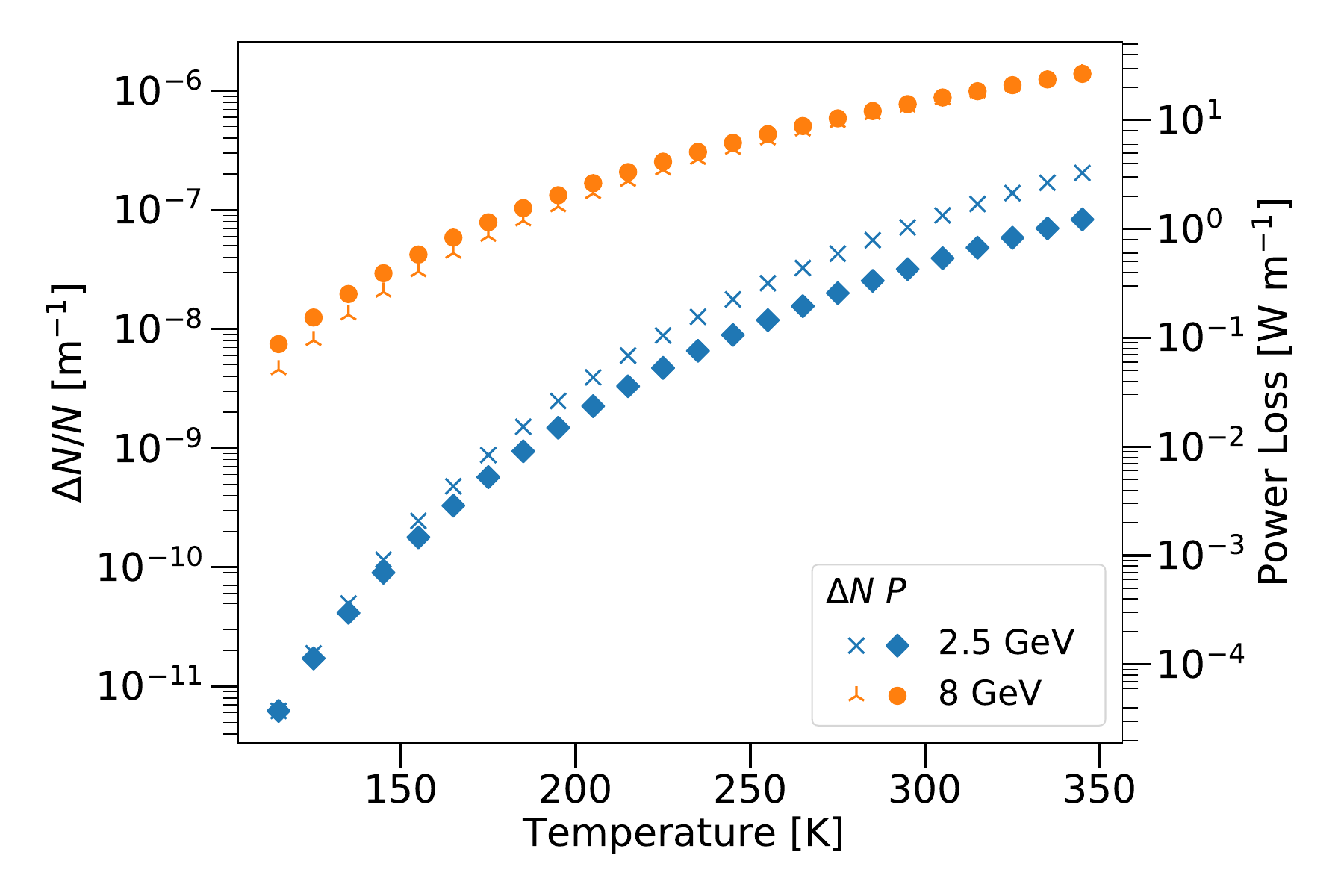}
  \vspace*{-8 mm}
  \caption{Blackbody-radiation stripping for 60~mA beams at a 4\% duty cycle for 2.5 and 8~GeV. Markers in the left and right-hand columns of the legend correspond to fractional loss rate and corresponding beam power loss, respectively.}
  \label{blackbody}
\end{figure}
This fractional loss per unit length can be modeled as~\cite{carneiro_beam_nodate,carneiro_numerical_2009,bryant_atomic_2006, du_photodet_1988}

\begin{align}
    &\frac{\Delta N}{N}\frac{1}{L} 
    =\int_{0}^\infty d\epsilon \int_0^\pi d\alpha \frac{d^3 r}{d\Omega d\nu d l} 
    \\ \nonumber
    &\frac{d^3 r}{d\Omega d\nu d l} =\frac{\left(1+\beta cos\alpha \right)n(\nu,r)\sigma(v')}{4\pi\beta}~~~,
\end{align}

\noindent where $n(\nu,r)$ is the spectral density of thermal photons as a function of frequency $\nu$ and radius $r$; $\alpha$ is the angle between the incoming photons and the beam; $\epsilon=h\nu/E_0$, with $E_{0}=0.7543~\mathrm{eV}$ being the electron binding energy for $H^-$; and $h$ is Planck's constant. The stripping cross section in the beam frame,  $\sigma(\nu')$, is given empirically by~\cite{armstrong_empirical_1963}

\begin{equation*}
    \sigma(E') = 8\sigma_{max_{BB}}
    E_0^{3/2}\frac{
    \left(E'-E_0\right)^{3/2}
    } {E'^3}~~~,
    \label{BB_xsect}
\end{equation*}
where
$\sigma_{max_{BB}}=4.2 \times 10^{-17} \mathrm{~cm}^{2}$ is the maximum blackbody-stripping cross section. These equations can be evaluated as

\begin{align}
 &\frac{\Delta N}{N}\frac{1}{L} = 
 \frac{ 8~\sigma_{max_{BB}}~E_0^{3/2} } {2\pi^2 \beta \gamma^3 (\hbar c)^3}
 \int_0^\infty dE'~~\mathlarger{{\times}} 
 \nonumber
 \\
 &\int_{-1}^{+1} du \frac{1}{(1+\beta u)^2}\frac{(E'-E_0)^{3/2}}{E'}\left[\thinspace e^{\mbox{\large $ \left(\large{\frac{E'}{kT\gamma(1+\beta u)}}\right)$}}-1\right]^{-1},
\label{BB_integral}
\end{align}
 
\noindent where $\gamma$ is the relativistic Lorentz parameter, and $kT$ is the Boltzmann constant times absolute temperature. The integral over $du$ can be performed analytically, then the outer integral can be calculated numerically \cite{carnier_beamsdoc,bryant_atomic_2006}. This analysis ignores the radial dependence in the photon density, that is $n(\nu,r)\approx n(\nu)$, this dependence is discussed shortly. Calculations using this approach for pulsed 60~mA beams are shown in Fig.~\ref{blackbody}. Here, we see that the power loss due to blackbody stripping exceeds $0.1~\mathrm{W/m}$ for a 2.5~GeV beam at temperatures above 250 K; for room temperature, it reaches about  $0.5~\mathrm{W/m}$. (The typical activation limit allowing for timely hands-on maintenance in a proton accelerator is 1~W/m \cite{Mohkov:2000ue}.) A more sophisticated treatment, accounting for external fields and incoming photon polarization, was performed by Herling and Bryant \cite{herling_blackbody_2009}.

\begin{figure}[ht!]
  \includegraphics[trim=0 0 0 0,clip,width=0.45\textwidth]{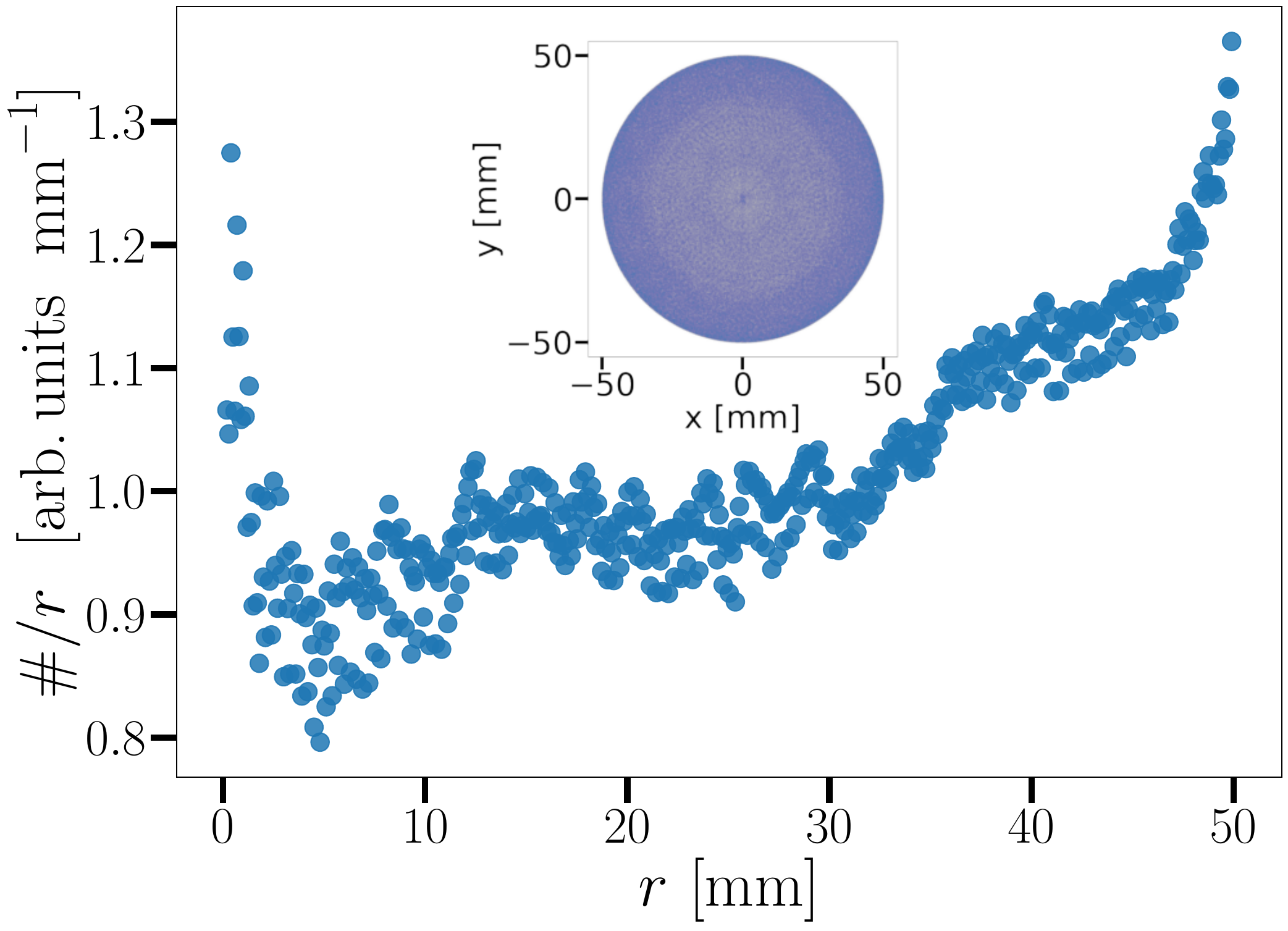}
  \vspace*{-3 mm}
  \caption{Radial $r$ dependence of blackbody radiation photon counts (normalised by $1  / r $) in a circular pipe. A reflectivity of 97\% is used, with a random noise factor added to each reflection to simulate surface roughness (with a maximum of $\pm 0.5^\circ$). Shown is a single-particle trajectory over 200,000 reflections with a 0.5~ps timestep, with absorption triggering immediate re-emission from a random point on the pipe diameter. Photon emission angles are generated using a beta distribution with $\alpha=1.65$, $\beta=1.65$ from $-\pi / 2$ to $\pi / 2$ radians.  Inset shows the same trajectory in a plane transverse to the beam axis.
  }
  \label{BB_radial}
\end{figure}

One may also consider the radial dependence of the photon density (i.e. inserting a radially dependent term to Eq.~[\ref{BB_integral}]). This dependence is illustrated in Fig.~\ref{BB_radial}. We note that reducing the maximum random-error angle at the reflection interface (shown is 0.5$^\circ$) results in a greater concentration of photons near the origin. We assume here that the photon emission angle distribution follows a beta distribution~\cite{gupta2004handbook}, with maximum emission perpendicular to the surface and a smooth fall-off to zero for emission tangential to the surface. This dependence is used in calculating the blackbody results shown in the cumulative stripping analysis in Section~\ref{beam_params}, but only has an effect of ${\lesssim}2\%$ on the overall stripping power loss versus a uniform photon distribution. More thorough analyses are available in~\cite{PROKHOROV2009181,wang_comparison_2013,nguyen_quang_evaluation_2017,ono_calculation_1980}.

Although blackbody stripping is not an issue in low-temperature regions of a linac (especially superconducting linacs), transfer lines between the accelerating linac sectors and the accumulator ring are not necessarily cooled. For laboratories with existing cooling infrastructure, or for transfer lines shorter than a few tens of meters, bringing such sections to 100--200~K may be a cost-effective means of limiting blackbody stripping.

If this is not feasible, low-emissivity beam-pipe coatings may be a worthwhile alternative. To our knowledge, these have not been studied in the context of blackbody stripping. Emissivity depends on surface conditions such as roughness and oxidation: stainless steel may have an emissivity at room temperature of up to 0.4, but this value can be as low as 0.03 for a well polished metal such as Cu, Ni, or Au~\cite{handbook-thermal-imaging, HUANG20106893, SETIENFERNANDEZ2014549}. As a candidate material, TiN is interesting, owing to its mechanical and thermal properties~\cite{handbook-infrared, YUSTE20111784,ZHAO20081272}; but other alternatives could also be considered \cite{chiba_low-emissivity_2005, hwang_high_2019}.

\subsection{\label{LZST}Lorentz Stripping}
In the beam frame, the transverse external magnetic field of focusing and bending elements Lorentz transforms into an electrical one\footnote{For a general example, see \cite{jackson_classical_1999}}
\begin{equation}
    \label{lor_E_B}
    |E_\perp| = \beta\gamma c |B_\perp|~~~.
\end{equation}

\noindent As a consequence, H$^-$ beams reaching the GeV energy range can have problematic stripping from magnetic optics at the field strengths required for steering and focusing. 

\begin{figure}
  \includegraphics[width=0.48\textwidth]{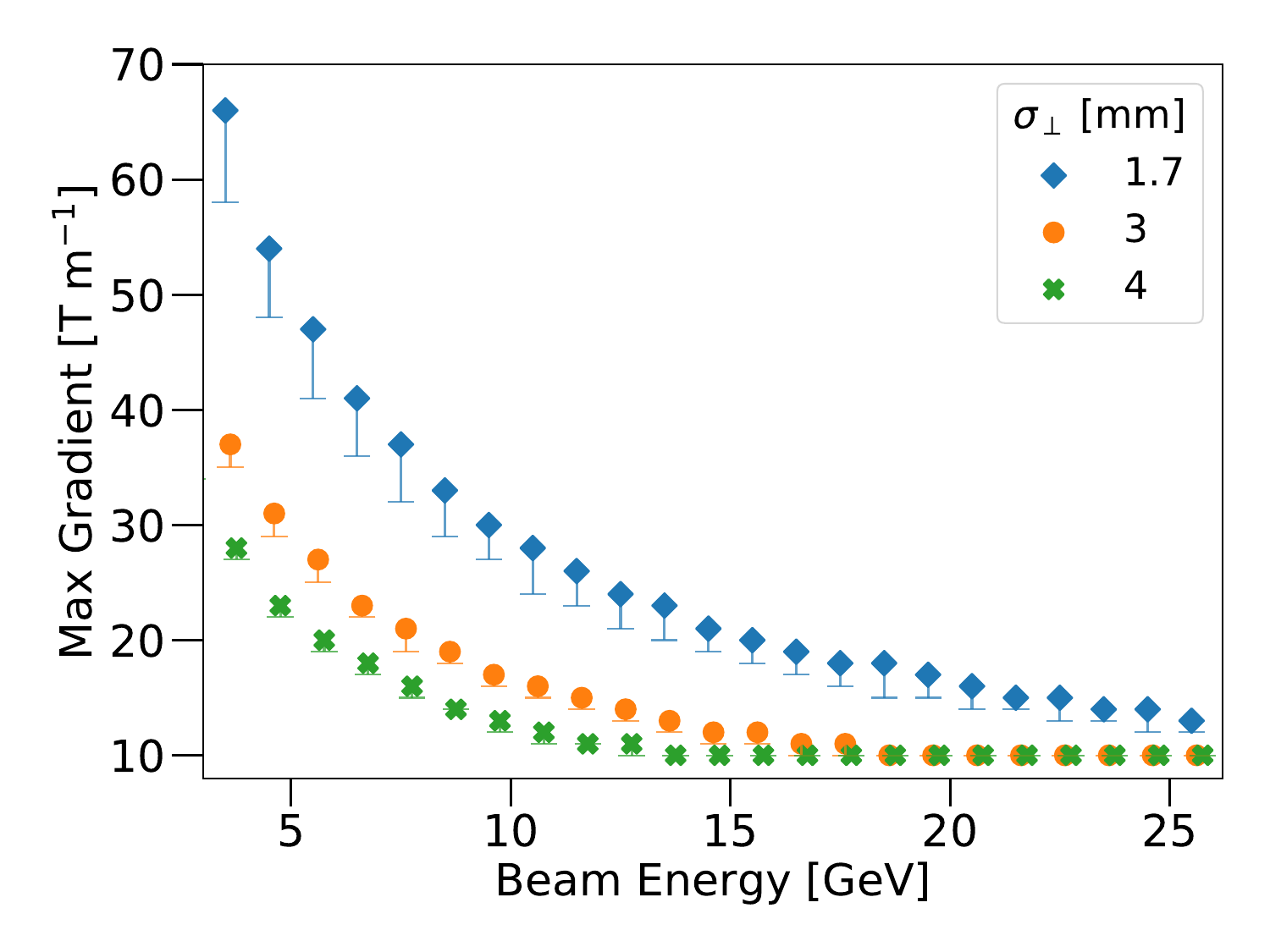}
  \vspace*{-8 mm}
  \caption{
  Maximum quadrupole field gradients for maintaining a Lorentz stripping beam-power loss rate of less than 0.1~W/m as a function of beam energy in GeV. A radial Gaussian distribution is used, with error bars reflecting a $\mu=0.7$~mm offset, see Eqs.~(\ref{lor_quad}) and (\ref{lorentz_integral2}).}
  \label{lorentz}
\end{figure}

The stripping probability per unit length for this effect can be modeled as \cite{keating_electric-field_1995,jason_neutralization_1981}
 \begin{align}
    \frac{\Delta N}{N}&\frac{1}{L} = \frac{|B_{\perp}|}{A_1}e^{-\frac{A_2}{\beta \gamma c \left|B_{\perp}\right|}}
    \\ \nonumber
    A_1 &= 3.073 \times 10^{-6}~\mathrm{s~V / m}
    \\ \nonumber 
    A_2 &= 4.414 \times 10^9~\mathrm{V / m}~~~.
    \end{align}

where $A_1$ and $A_2$ are empirical-fit parameters. With this equation, calculating Lorentz stripping for bunches traversing dipole magnet fields is straightforward. Quadrupoles, however, have a transverse field-strength dependence which makes particles in the outer halo or misaligned beams more likely to undergo stripping. This makes smaller rms transverse beam size $\sigma_{\perp}$ advantageous (see Fig.~\ref{lorentz}). However, the opposite is true for IBSt (see Sec.~\ref{IBST}). 

This stripping probability through a quadrupole can be modeled as

\begin{equation}
P = \int^{2\pi}_{0}\int^r_0 f(r',\sigma)\frac{\Delta N}{N} \frac{1}{L} r' dr d\theta~~~,
\label{lor_quad}
\end{equation}

\noindent where $L$ is the quadrupole length and $f(r',\sigma)$ describes a particle distribution \cite{posocco_magnetic_nodate}. 
Since magnet or beam misalignments can substantially increase Lorentz stripping, stricter error tolerances should be expected than would otherwise be necessary \cite{neven_2021}. A more detailed analysis of Lorentz stripping in terms of beam parameters (and in relation to IBSt) is given in Section~\ref{beam_params}. 

\subsection{\label{IBST} Intrabeam Stripping}
The tendency of H$^-$ ions to collide within a bunch can be the prevalent stripping mechanism in modern high intensity H$^-$ accelerators \cite{shishlow_intra_2012}. For this effect, the loss rate per unit length along the beam axis in the laboratory frame can be modeled as 

\begin{equation}
\frac{\Delta N}{N}\frac{1}{L} = 
\frac{N\sigma_{max_{IB}}
\sqrt{\gamma^2\theta_x^2+\gamma^2\theta_y^2+\theta_z^2}
}%
{8\pi^2\gamma^2\sigma_x\sigma_y\sigma_z}F\left({\gamma\theta_x,\gamma\theta_y,\theta_z}\right),
\label{ibs_lossperm}
\end{equation}
\\
\noindent  where $F\left({\gamma\theta_x,\gamma\theta_y,\theta_z}\right)$ is a shape function for bunch-frame momentum spreads (the latter are defined below). This function can be approximated as

\begin{equation}
F(a, b, c) \approx 1+\frac{2-\sqrt{3}}{\sqrt{3}(\sqrt{3}-1)}\left(\frac{a+b+c}{\sqrt{a^{2}+b^{2}+c^{2}}}-1\right),
\end{equation}
and varies with weak dependence on its parameters from 1 to 1.15. (See \cite{lebedev2012intrabeam} for a detailed treatment, or \cite{cohen_stripping_1986} for a discussion of the underlying physics.) Then, $\sigma_{max_{IB}}$ is the maximum IBSt cross-section (not to be confused with $\sigma_{max_{BB}}$ from Eq.~[\ref{BB_xsect}]). This can be estimated as

\begin{equation}
\nonumber
\sigma_{max_{IB}} \approx \frac{240 a_{0}^{2} \alpha_{f}^{2} \ln{\left(1.97\frac{\alpha_{f} + \beta}{\alpha_{f}} \right)}}{\left(\alpha_{f} + \beta\right)^{2} }~~\lesssim~~4\times 10^{-15}~\mathrm{cm}^{-2},
\end{equation}

\noindent with the fine-structure constant $\alpha_f$, the Bohr radius $a_0$, and with $\beta=v / c$ in this case being the relativistic velocity between ions. The rms bunch sizes and angular momentum spreads are then defined as
\begin{equation}
\nonumber
\sigma_{x,y} = \sqrt{\epsilon_{x,y}\beta_{x,y}}
    \quad\quad
    \theta_{x,y} = \sqrt{\frac{\epsilon_{x,y}}{\beta_{x,y}}}~~.
\end{equation}

\begin{figure}
  \includegraphics[width=0.48\textwidth]{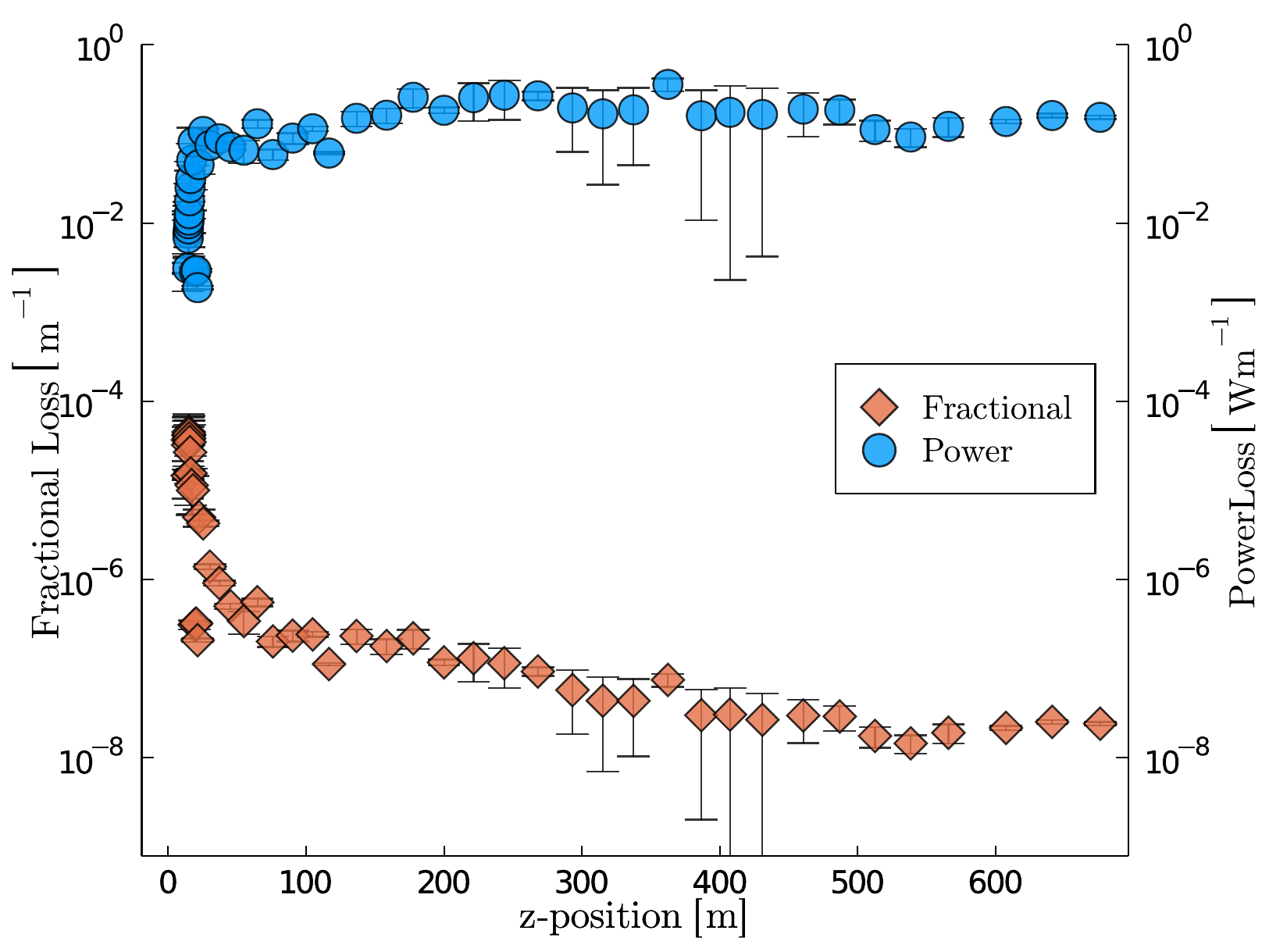}
  \vspace*{-8 mm}
  \caption{
  Stripping losses, as per Eq.~(\ref{ibs_lossperm}), and corresponding power loss rate along a pulsed 62.5{\thinspace}mA H$^-$ beam in a superconducting linac accelerating to 2.5~GeV with a 4\% duty cycle (nominal ESS / ESS$\nu$SB upgrade design~\cite{Garoby_2017}).}
  \label{loss_pow}
\end{figure}

\begin{figure}[ht!]
  \includegraphics[trim=10 0 101 0,clip,width=0.5\textwidth]{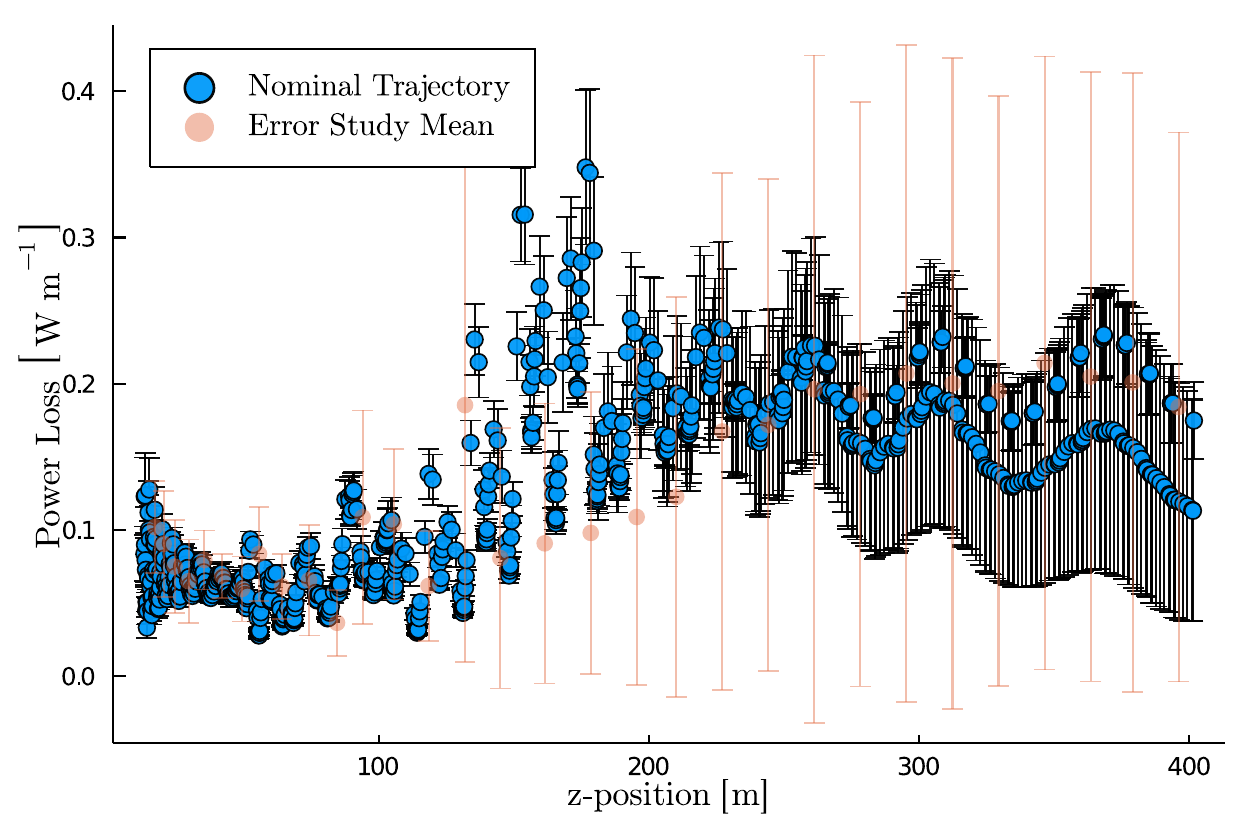}
  \vspace*{-8 mm}
  \caption{IBSt Power losses for a nominal trajectory and a corresponding error study for an H$^-$ linac identical to that in Fig.~\ref{loss_pow}. Both static and dynamic machine errors and beam misalignment were incorporated, with standard deviations taken from the cumulative trajectories of 100 trials.}
  \label{power_errstud}
\end{figure}

\noindent where $\epsilon_{x,y}$ and $\beta_{x,y}$ are the transverse Twiss parameters.

The inverse dependence of Eq.~(\ref{ibs_lossperm}) on the longitudinal bunch size $\sigma_z$ means that IBSt can be limited by maximizing the bunch length or, similarly, by minimizing momentum spread.

Additional dependence on accelerating phase and RF frequency may also be inferred, either of which can be reduced to mitigate stripping. Relaxed transverse focusing also limits IBSt; in this case, one can take the defocusing strength of cavities as a limiting parameter for a minimally focused beam~\cite{lebedev2012intrabeam}.

Figure~\ref{loss_pow} shows the stripping loss rate per unit length and corresponding power loss for a 5~MW, 2.5~GeV H$^-$ linac. These results were calculated as a post-processing step using trajectories from TraceWin~\cite{uriot2011tracewin}. We adopted this approach since stripping loss rates are typically orders of magnitude below that of a bunch population (despite causing problematic activation), and thus unlikely to affect bunch dynamics\footnote{With the exception of residual gas stripping in the LEBT, which can eliminate a large proportion of an H$^-$ beam. TraceWin does have a built-in functionality for calculating gas stripping and conventional dynamics in tandem.}. Figure~\ref{power_errstud} then compares power losses with error study trajectories along the same linac as Fig.~\ref{loss_pow}. The error study parameters include static and dynamic dipole, quadrupole, and cavity errors, as well as beam misalignments.\footnote{These included 0.1~mm and 10~mrad static errors for the beam distributions position and momentum coordinates, respectively; 0.1~mm displacement, 1$^\circ$ rotation and 0.5\% gradient errors for quadrupoles; and 1\% field and 0.5\% phase error for cavities.}

\subsubsection{H$^0$ Traversal and Power Deposition}
\begin{figure}[t!]
  \includegraphics[width=0.5\textwidth]{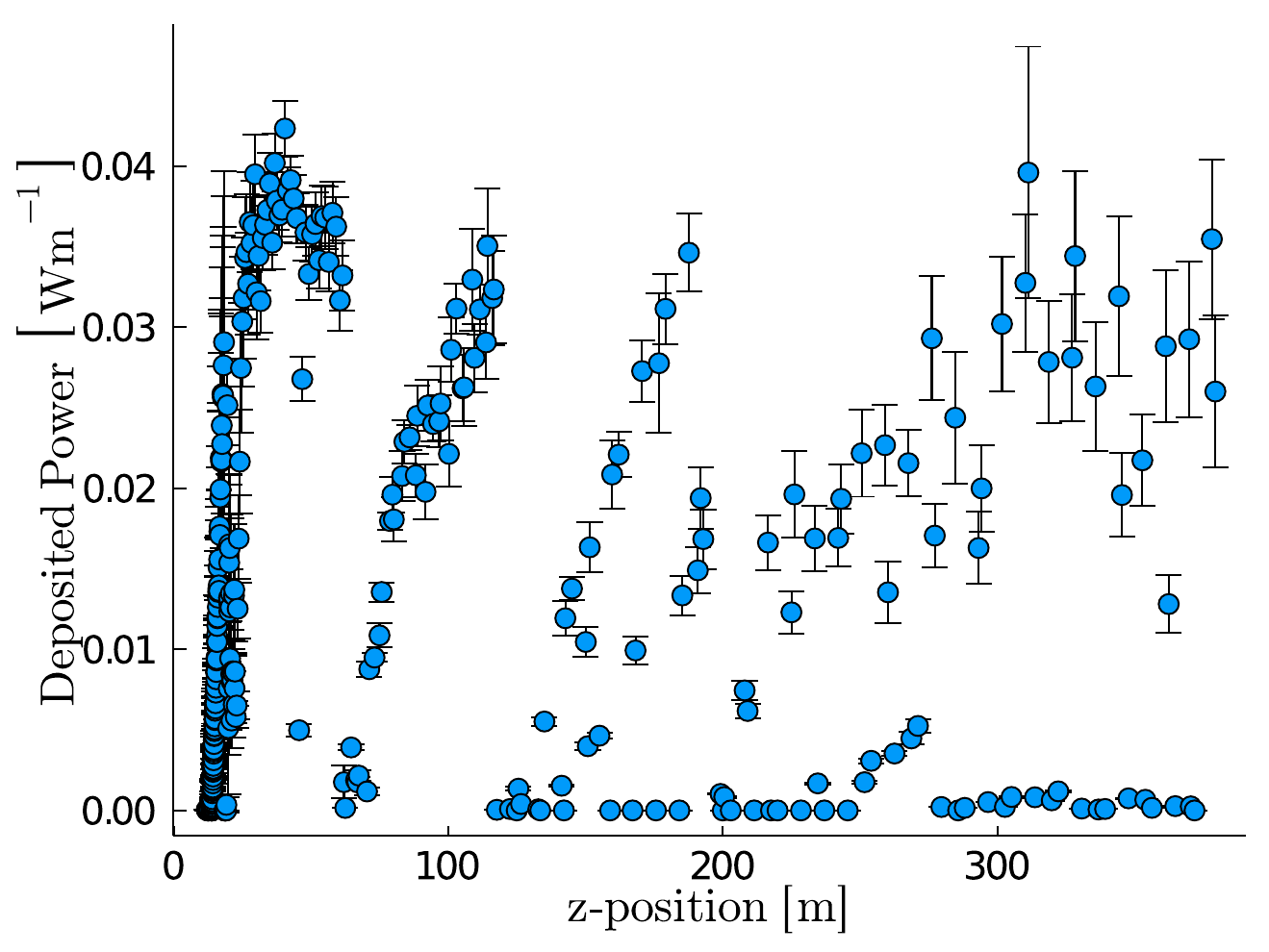}
  \vspace*{-8 mm}
  \caption{Power deposition of H$^0$ into beam-pipe and machine element walls for a 62.5{\thinspace}mA H$^-$ beam accelerating to 2.5~GeV as in Fig.~\ref{loss_pow}. Only IBSt is accounted for.}
  \label{power_dep}
\end{figure}

In Fig.~\ref{power_dep}, the H$^0$ particles generated by IBSt are tracked from the point of stripping to collision with the beam pipe or other machine element. This calculation was performed using an elementary trajectory integrator, since the neutral H$^0$ particles do not interact with the machine optics or cavity fields.

Error studies were also performed with this parameter set  using the same approach as shown in Fig~\ref{power_errstud}. These error study runs (not shown) resulted in a modest increase of deposited power to an average of roughly 0.05~W/m. One can compare this with Fig.~\ref{power_errstud}, where instantaneous beam-power loss (stripped, but not tracked to a collision point) reaches 0.2~W/m through the high-energy sectors.

\begin{figure}[h!]
  \includegraphics[width=0.48\textwidth]{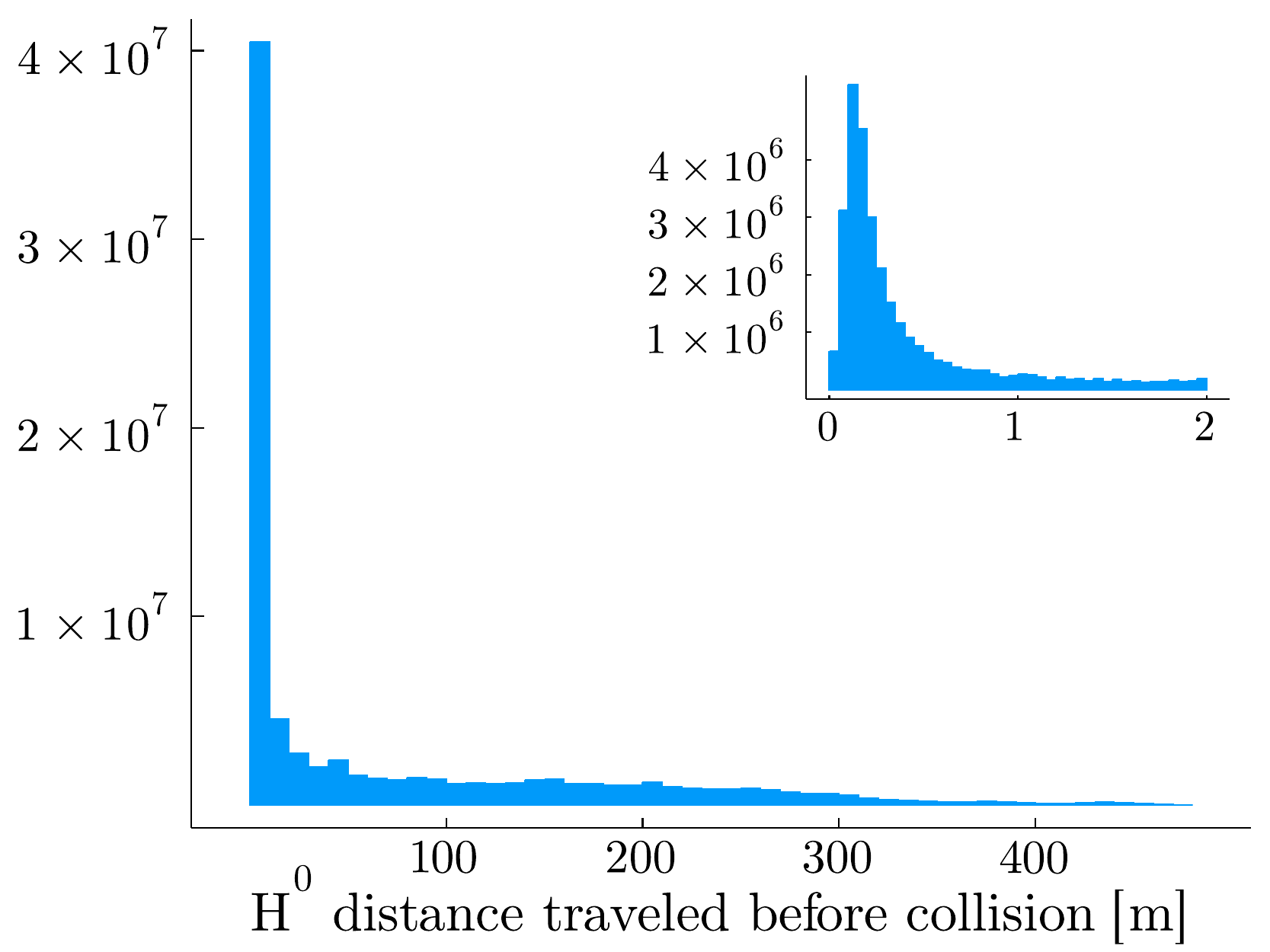}
  \vspace*{-6 mm}
  \caption{Traversal of H$^0$ particles from the point of IBSt to collision with the nearest machine-element wall or aperture (for the same beam and machine parameters used in Fig.~\ref{power_dep}). Inset shows the depositions occurring within 2~m of stripping.
  }
  \label{H0distance}
\end{figure}

This disparity between raw power loss and power deposited in-flight is a notable improvement, especially at energies exceeding 2~GeV, and recalling that the practical activation limit for machine maintenance is 1~W/m. To clarify this concept, Figure~\ref{H0distance} illustrates the distances traveled by the stripped H$^0$ particles of a bunch (with a remaining $\sim$50~W deposited into the dump). Note that we cannot expect this advantage in a transfer line: the H$^0$ traversal distance is limited by the dipole bends, which can be expected to create activation hotspots where miniature dumps may be advisable.

\section{Design Recommendations}
\subsection{\label{beam_params} Optics, Emittance, and Phase Advance}
For machine sections requiring bending magnets, the consequent Lorentz stripping scales with the bending radius (e.g. $\rho=73$~m for a field strength limited to $|B|=0.15$~T in a 5~MW, 2.5~GeV beam \cite{neven_2021}). This sets a strong engineering constraint, especially for projects with space limitations, which may need to adopt more stringent measures for limiting residual-gas and blackbody stripping (higher vacuum and lower temperature, respectively). From a design perspective, as beam energies reach $\sim$8~GeV, Lorentz stripping can become problematic with magnet strengths as low as 50~mT \cite{herling_blackbody_2009}. This makes higher energies effectively untenable for H$^-$, although designing for a large longitudinal bunch size, large machine aperture, and low quadrupole gradients may extend this limit.  

As mentioned in Secs.~\ref{LZST} and \ref{IBST}, relaxed quadrupole focusing can improve the IBSt rate while being detrimental for Lorentz stripping, as particles furthest from the beam axis and closer to the magnet pole tips encounter an increased gradient.

\begin{figure}[ht!]
  \includegraphics[width=0.5\textwidth]{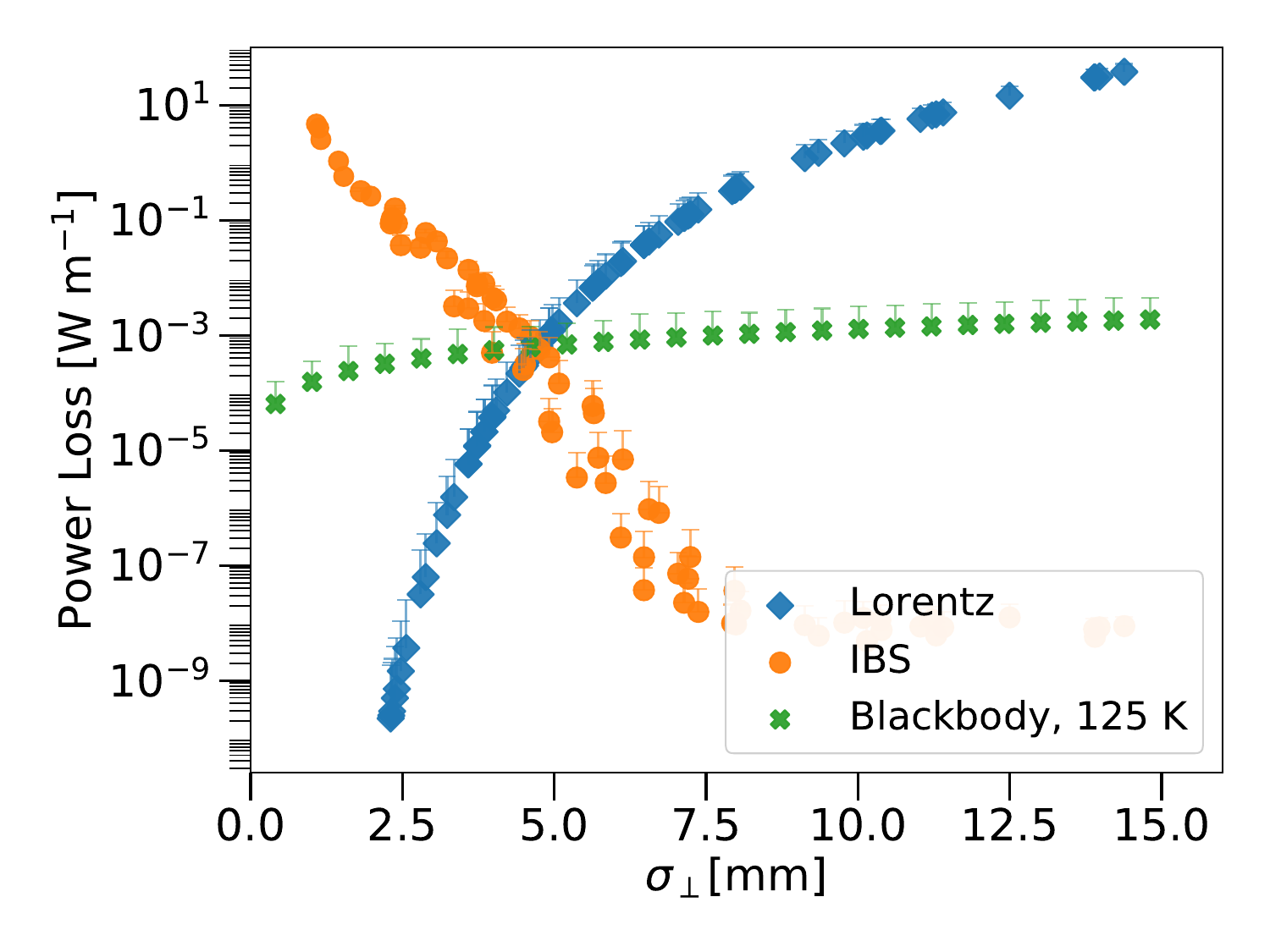}
  \vspace*{-8 mm}
  \caption{Dependence of blackbody, IBSt and Lorentz stripping (quadrupoles only) on average transverse beam size $\sigma_\perp$ for a 2.5~GeV beam traversing a FODO lattice (6${\thinspace}$m~$\times$~20 cells) of one quadrupole pair and one bunching gap per cell. Beam parameters are determined by setting phase advance and solving for optimum inputs. A range of 1--90$^\circ$ phase advance runs gives the resulting range of beam sizes. Blackbody stripping is simulated separately, assuming a constant $\sigma_\perp$ for each point.
  }
  \label{lor_v_ibs}
\end{figure}

\begin{figure}[ht!]
  \includegraphics[width=0.5\textwidth]{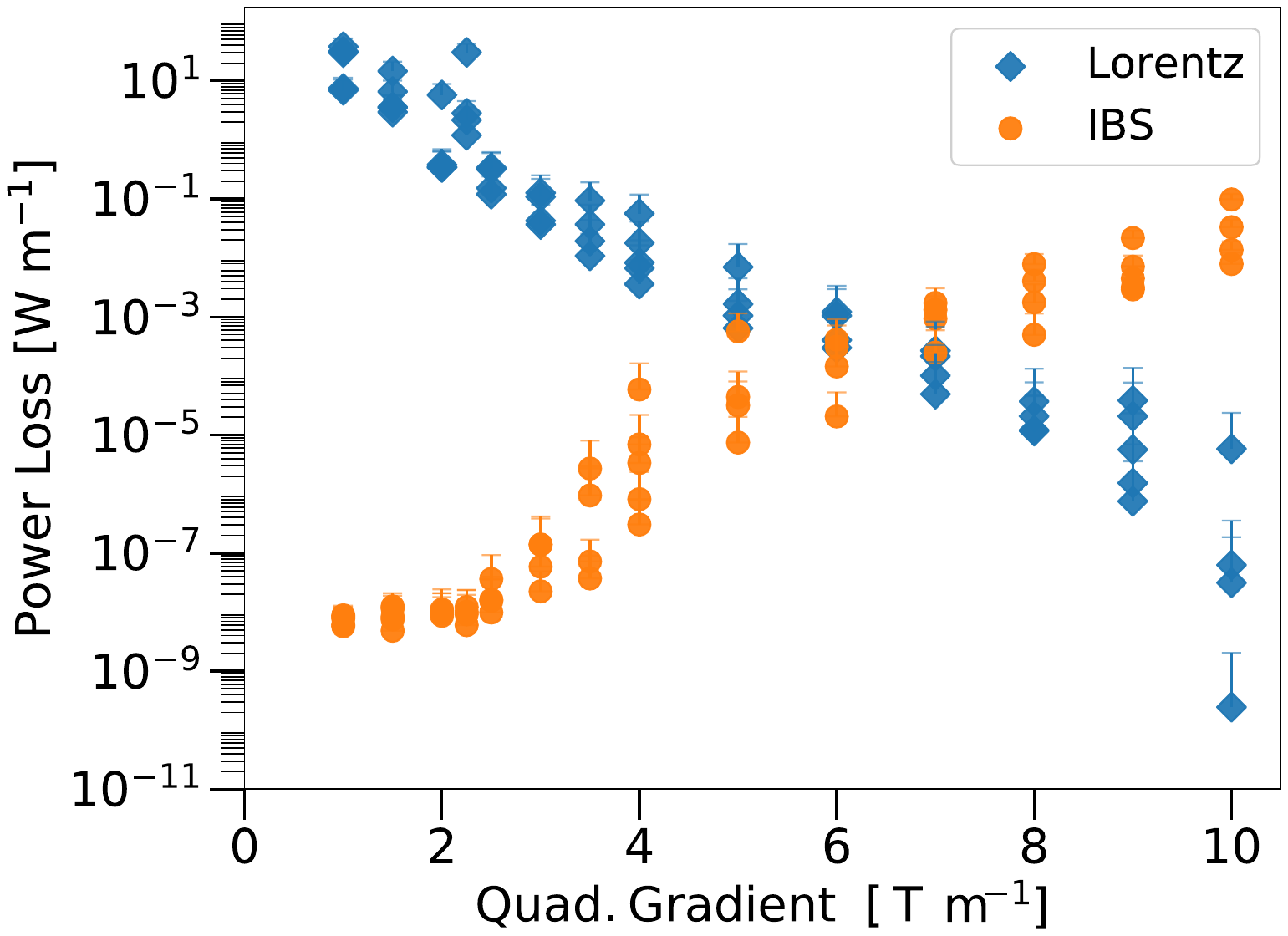}
  \vspace*{-8 mm}
  \caption{Dependence of IBSt and Lorentz stripping (quadrupoles only) on field gradient for the same lattice and optimization scheme as Fig.~\ref{lor_v_ibs}.  
  }
  \label{lor_v_IBS_quadgrad}
\end{figure}

We can also expand on this (and the discussion in \cite{lebedev2012intrabeam}) in terms of phase advance and emittance. Phase advance $\phi$ has the proportionalities
\begin{equation*}
\phi\sim\sigma_{\perp}^{-1/2}\sim\sqrt{|E|}~~~,  
\end{equation*}
where $\sigma_{\perp}$ and $|E|$ are the respective transverse beam size and focusing field strength (Lorentz transformed, as per Eq.~[\ref{lor_E_B}]) and $\sigma_{\perp}\cdot|E|$ is roughly constant. This means Lorentz stripping is affected dually by phase advance: an increased phase advance reduces $\sigma_{\perp}$, in turn reducing exposure to the high-field regions in the quadrupoles and limiting stripping. This dependence is illustrated in Fig.~\ref{lor_v_ibs}.\footnote{Meanwhile, a greater $|E|$ is required to increase the phase advance, which should increase stripping. However, we can see that the dependence on beam size dominates, with Fig.~\ref{lor_v_IBS_quadgrad} showing an inverse proportionality between Lorentz stripping and gradient strength.} For IBSt, the reduced $\sigma_{\perp}$ and increased $|E|$ associated with an increased phase advance are both detrimental. 

Figures~\ref{lor_v_ibs} and \ref{lor_v_IBS_quadgrad} illustrate this dependence for both stripping types. Here we use Eq.~(\ref{ibs_lossperm}) for IBSt, and Eq.~(\ref{lor_quad}) with a Gaussian distribution reduced to the following form for Lorentz stripping, which can be integrated numerically over a radius $r$:
\begin{equation}
    \frac{\Delta N}{N}\frac{1}{L} = 
    \int_{0}^{3 \sigma} \frac{ \sqrt{2} G r^{\prime}}{2 \sqrt{\pi} A_{1} \sigma}
    e^{-\frac{\left(-\mu+r^{\prime}\right)^{2}}{2\sigma^{2}}} e^{-\frac{A_{2}}{G \beta\gamma c}}
     d r~~~.
     \label{lorentz_integral2}
\end{equation}
A misalignment of $\mu = 1.5$~mm is used in Fig.~\ref{lor_v_ibs} to give the upper error; $G$ is the quadrupole field gradient in T/m. Additionally, longitudinal/transverse emittance ratios were varied from 0.5 to 2 at each phase-advance step, giving a realistic margin for expanding or contracting longitudinal bunch size to accommodate stripping limits. The blackbody-stripping results rely on Eq.~(\ref{BB_integral}), with the additional radial dependence as per Fig.~\ref{BB_radial} convolved to a Gaussian beam.

\begin{figure}[h!]
  \includegraphics[width=0.5\textwidth]{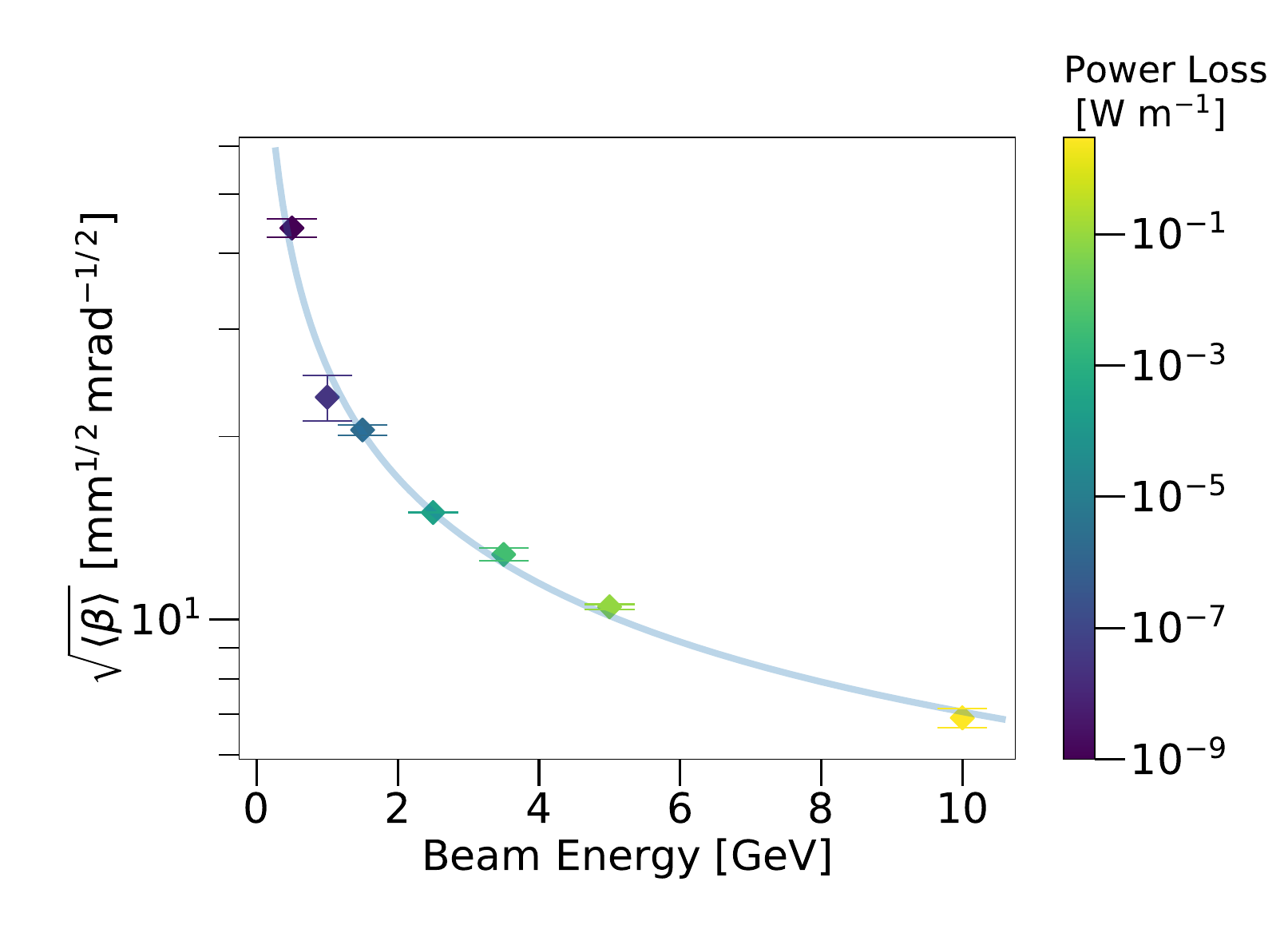}
  \vspace*{-8 mm}
  \caption{Points of intersecting Lorentz and intrabeam stripping levels as a function of beam energy and average transverse Twiss parameter $\left< \beta \right>$, using the same lattice structure and optimization routine as Fig.~\ref{lor_v_ibs}. The solid line shows an exponential fit reflecting Eq.~(\ref{bal_curvefit}). The color bar shows the power loss at each point (color online). The Twiss $\left< \beta \right>$ dependence is used as an emittance-independent measure of beam size for a smoother curve fit.}
  \label{bal_pts}
\end{figure}

For this test, separate runs were performed with varied quadrupole strengths and gap voltages to cover a phase-advance range of 1--90$^{\circ}$.
Figure~\ref{bal_pts} compiles the crossing points for Lorentz stripping and IBSt using the same lattice structures at varied energies. Here we use the average transverse Twiss parameter $\sqrt{\left<\beta\right>}$ as an emittance-independent measure of beam size. These can be considered optimal conditions for stripping-loss reduction in low-temperature, high-vacuum sectors (i.e. ignoring blackbody and gas stripping); with the curve fit allowing us to estimate the dependence of beam energy, $E_B$, on $\left<\beta\right>$:

\begin{equation}
     \sqrt{\left<\beta\right>} = 13.7~E_B^{-0.65}+1.65
    \label{bal_curvefit}
\end{equation}

For emittance, the situation is more straightforward. Increased emittance means larger $\sigma_{\perp}$, which increases Lorentz stripping; it also indicates greater velocities within a bunch, which increases IBSt. Thus, limiting emittance growth is especially important with H$^-$ beams. This characteristic was confirmed when comparing preliminary simulation studies for this work (high emittance growth) with the final, well-matched lattice (low emittance growth); the preliminary lattices with higher emittance growth had higher IBSt rates by a factor of 1.5${\sim}$2.

\subsection{\label{A2R} Test Case: Linac-to-accumulator transfer line}
Our calculations on H$^0$ power deposition afford some relief in terms of linac design, but this does not apply to transfer lines: here the stripped H$^0$ collide with bends in the transfer line instead of having a large fraction propagating to the end of the linac where they can be dumped.

Moreover, despite temperature and vacuum levels not necessarily needing to accommodate superconducting elements in a transfer line, activation from residual gas stripping can become problematic at a much stronger vacuum level than that needed to avoid gas scattering (roughly 0.001 and 0.1~Pa, respectively~\cite{ravelli_laurence_2019}). Meanwhile, blackbody-radiation stripping becomes problematic, as illustrated in Sec.~\ref{BBS}.

Lorentz stripping is also compounded in transfer lines, with bending dipoles making a non-negligible contribution, and with the focusing strengths of the quadrupoles no longer bound solely to the defocusing of the accelerating cavities.

These detriments imply that transfer lines may have to take into account a considerably larger activation level than their respective linacs. In other words, \textit{all} forms of stripping must be balanced when designing these sections, and the relevant cost estimates should include some moderate cooling (or beam-pipe coatings) and vacuum systems. Preliminary studies from our group indicate that relatively tight limits on focusing strength, phase advance, and emittance growth are necessary to keep Lorentz stripping and IBSt within acceptable activation limits for a 2.5~GeV, 5~MW beam~\cite{neven_2021}.
\\
\\
\section{Conclusions}
This paper summarizes the known H$^-$ stripping phenomena in the context of research-accelerator design. Although a number of the works cited herein provide superior analysis, we found that a comprehensive review was lacking. For a more complete study, the reader is strongly recommended to consult these works in particular: \cite{plum_beam_2016}, \cite{carnier_beamsdoc}, \cite{lebedev2012intrabeam}.

We have demonstrated that the proportion of stripped H$^0$ which traverse to the end of a linac, or local straight section, presents a significant fraction of the lost power from the beam (compared with the more problematic H$^0$ colliding with the beam pipe or machine-element walls). This may relax the limits for allowable stripping-based beam loss in H$^-$ accelerators. In general, we should stress the importance of performing trajectory analysis for stripped particles -- this was also shown to be critical for double-stripped particles in~\cite{ikegami_2012_beamcommiss}.

The balance points between Lorentz stripping and IBSt seen in Figs.~\ref{lor_v_ibs}, \ref{lor_v_IBS_quadgrad}, and \ref{bal_pts}, and modeled in Eq.~(\ref{bal_curvefit}), may be of general use for H$^-$ design studies: while specific lattice stripping rates may vary, these optima provide a baseline for determining feasible beam sizes.

Although the severity of the types of stripping as listed in the introduction agrees with our analysis, none are negligible in the design of a H$^-$ linac. Owing to this, care should be taken from the design stage to predict relevant types of stripping in each sector and ensure that they are kept below reasonable limits.

\vspace{0.5cm}
\section{Acknowledgements}
The authors would like to thank Dr. Ryoichi Miyamoto for his insights in producing this work.

This project has received funding from the European Unions Horizon 2020 research and innovation program under grant agreement No 777419.

\bibliography{biblio}

\end{document}